\documentclass[traditabstract]{aa}
\usepackage{txfonts}
\usepackage{natbib}
\usepackage{supertabular}

\newcommand{\kms}{km~s${}^{-1}$}

\begin{document}

\title{Radio Recombination Lines at Decametre Wavelengths}
\subtitle{Prospects for the Future}

\author{%
  W.~M.~Peters\inst{\ref{inst:NRL}} \and 
  T.~Joseph~W.~Lazio\inst{\ref{inst:NRL},\ref{inst:NLSI},\ref{inst:JPL}} \and 
  T.~E.~Clarke\inst{\ref{inst:NRL}} \and 
  W.~C.~Erickson\inst{\ref{inst:UTaz}} \and 
  N.~E.~Kassim\inst{\ref{inst:NRL}}
}

\institute{%
  Naval Research Laboratory, 4555 Overlook Ave.~\hbox{SW}, Washington,
	DC  20375 USA\ \email{Wendy.Peters@nrl.navy.mil,Tracy.Clarke@nrl.navy.mil,Namir.Kassim@nrl.navy.mil}\label{inst:NRL}
\and
  NASA Lunar Science Institute, NASA Ames Research Center, Moffett
	Field, CA  USA\label{inst:NLSI}
\and
  current address: Jet Propulsion Laboratory, M/S 138-308, 4800 Oak
Grove Dr., Pasadena, CA  91109, USA\ \email{Joseph.Lazio@jpl.nasa.gov}\label{inst:JPL}
\and
  University of Tasmania, Churchill Ave., Sandy Bay, Tas 7005 Australia\label{inst:UTaz}
}

\date{%
  Received: 2010 /
  Accepted: 2010 August~30
}

\abstract{
{This paper considers the suitability of a number of emerging
and future instruments for the study of
radio recombination lines (RRLs) at frequencies below~200~MHz.}
These lines are of interest because they arise only in low-density
regions of the ionized interstellar medium and because they may
represent a frequency-dependent foreground for next-generation
experiments trying to detect \ion{H}{i} signals from the Universe's
Epoch of Reionization and Dark Ages (so-called ``21-cm cosmology''
observations).
{We summarize existing decametre-wavelength observations of
RRLs, which have detected only carbon RRLs.}
We then show that, with reference to an interferometric array, the
primary instrumental factor limiting detection and study of the RRLs
is the areal filling factor of the array.  
{We consider the first station of the Long Wavelength Array
(LWA-1), the LOw Frequency ARray (LOFAR), the low-frequency component
of the Square Kilometre Array (SKA-lo), and a future Lunar Radio Array
(LRA), all of which are likely to operate at decametre wavelengths.}
{Key advantages that many of these arrays offer include digital
  signal processing, which should produce more stable and better
  defined spectral bandpasses; larger frequency tuning ranges; and better angular resolution than that of the previous
generation of instruments that have been used in the past for RRL
observations.}
Detecting Galactic carbon RRLs, with optical depths at the level of
$10^{-3}$, appears feasible for all of these arrays, with integration
times ranging from a few hours to as much as 100~hr; at optimal
frequencies this would permit a Galactic survey.  The SKA-lo and
\hbox{LRA}, and the LWA-1 and LOFAR at the lowest frequencies, should
have a high enough filling factor to detect lines with much lower
optical depths, of order $10^{-4}$ in a few hundred hours.
{The amount of RRL-hosting gas present in the Galaxy at the high
Galactic latitudes likely to be targeted in Epoch of Reionization and
Dark Ages \ion{H}{i} studies is currently unknown.}
{If present, however, the spectral fluctuations from RRLs could
be comparable to or exceed the anticipated HI signals.}
}


\maketitle

\section{Introduction}\label{sec:intro}

In the low-density regions of the Galactic interstellar medium (ISM),
free electrons can be captured by ions at very high quantum numbers
($n > 100$).
{As the atom cascades into a series of successively lower
ionization states, each transition produces a radio recombination line
\citep[RRL,][]{gs02}.}
These Rydberg atoms are large, often having macroscopic dimensions,
and are consequently somewhat fragile.
{As a result, the transitions between high quantum number
levels are a sensitive probe of the environments in which the atoms
exist, providing diagnostics such as temperature, density, and
pressure, and the lines can be used to constrain the size of the
regions in which the atoms occur.}
RRLs also may be a frequency-dependent contaminant in future
astrophysical and cosmological observations using the highly
redshifted \ion{H}{i} 21-cm line from the Epoch of Reionization (EoR)
or earlier.

\cite{1980Natur.283..360K} discovered the first decametre RRL at a
frequency of~26.13~MHz, using the UTR-2 telescope to observe
\object{Cas~A}.  \cite{1980Natur.287..707B} subsequently
identified the line as the C631$\alpha$ line.  Since then, carbon RRLs
have been measured towards \object{Cas~A} at frequencies from~14
to~1420~MHz \citep[and references therein]{1994ApJ...430..690P}.  A
recent study by \cite{2007MNRAS.374..852S} measured $\alpha$, $\beta$,
$\gamma$, and even $\delta$ carbon-line transitions at frequencies
near~20~MHz.  The lines were measured up to $n \sim 1005$, which is
the highest $n$-bound state ever detected.

Observationally, the RRLs resulting from high Rydberg transitions
typically occur in emission at frequencies above~200~MHz (shortward of
1.5~m) and have line widths dominated by Doppler broadening.
Below~150~MHz, however, the excitation temperature of the atoms
approaches the typical gas kinetic temperature and the lines appear in
absorption \citep{1989ApJ...341..890P}.  In between is a poorly
studied transition region, in which the lines may be undetectable.

{
There are a number of new telescopes, either under construction or in the
design phase (Sect.~\ref{sec:future}), that will be operating at frequencies below
about~200~MHz, which we shall henceforth consider ``decametre
wavelengths.''  This paper considers the
impact these emerging instruments may have on RRL studies,
specifically the $\alpha$-transition Carbon RRLs.  We review aspects
of decametric RRLs from the perspective of the Galactic interstellar
medium (Sect.~\ref{sec:ism}) and cosmological observations
(Sect.~\ref{sec:cosmo}), then in Sect.~\ref{sec:detect} we consider 
the detection of the lines, in Sect.~\ref{sec:future} we discuss the
specifics of various instruments with respect to RRL observations,
and, in Sect.~\ref{sec:conclude} we present our conclusions.  }

\section{RRLs and the Galactic ISM}\label{sec:ism}

Pressure and radiation broadening as well as Doppler effects may
affect the RRL absorption line profiles, and can be used to derive
physical properties of the absorbing cloud(s)
\citep{1998ApJ...506..758K}.  The relative strengths of the $\alpha$,
$\beta$, and~$\gamma$ lines, which can be measured simultaneously at
different quantum numbers~$n$ in the same spectrum, can be used to
constrain the physical properties of the absorbing gas further
\citep{1995ApJ...454..125E}.  While hydrogen, helium, and carbon lines
have been identified in emission, all low-frequency RRLs known to date
are high-$n$ carbon transitions.

Using multi-frequency model fits to all existing data on the
\object{Cas~A} sightline and comparing the spatial distribution of
the higher frequency lines to that of \ion{H}{i} and molecular clouds
on the sightline, \cite{1998ApJ...506..758K} concluded that these RRLs
most likely occur in diffuse \ion{H}{i} clouds in the Perseus
arm, which have temperatures of order 75~\hbox{K}.
The best fit physical models, based on \object{Cas A} sightline
data from all frequencies, predict a stronger broadening in the lowest
frequency lines than has been actually observed.  According to
\cite{1998ApJ...506..758K}, the discrepancies arise, at least in part,
from difficulties in accurately detecting and measuring the Lorentzian
wings due to the weakness of the lines and issues with accurate
spectral baseline removal.  Many lines reported in the literature have
been erroneously fit with Gaussians
\citep[e.g.,][]{1990ASSL..163..189S}, and others were subjected to a
baseline removal process which significantly diminished the fitted
line strengths and widths \citep{1994ApJ...430..690P}.  In order to
fully realize the potential of these lines in measuring physical
conditions in the \hbox{ISM}, sensitive observations with robust
spectral baselines are needed, particularly at the lowest frequencies.
With their very high signal-to-noise observations,
\cite{2007MNRAS.374..852S} argue that errors in fitting or spectral
baseline removal alone cannot account for the line-width discrepancy;
however including the known low-frequency spectral turnover of
\object{Cas~A} \citep[e.g.,][]{hk09} lowers the amount of
predicted broadening and brings the model and observations into
agreement.

{Aside from the \object{Cas~A} sightline, carbon recombination
lines in absorption have been observed on a variety of
sightlines in the Galaxy (Tables~\ref{tab:knownlines}
and~\ref{tab:previous}).}
Most observations have focussed on sightlines towards known bright
background sources or passing through gas-rich regions along the inner
Galactic plane, but the detection rates have been low.  Only two
broader surveys for the lines have been published.
\cite{1995ApJ...454..125E} found carbon RRL absorption complexes on
roughly 30 sightlines between Galactic longitudes of~340\degr\
and~20\degr, with Galactic latitudes $|b| < 2\degr$.  They concluded that there is no
evidence for pressure broadening in their data, and that the
linewidths reflect Doppler broadening and the presence of many clouds
in their 4\degr\ beam.  \cite{2001JApA...22...51K} found RRLs on~9
sightlines, with six of those in the Galactic longitude range
$352\degr < \ell < 17\degr$.  Combining their data with that of
\cite{1995ApJ...454..125E} and information from observations on the
same sightlines at higher frequencies, they conclude that the clouds
likely have sizes of~2\degr--4\degr\ and occur in photodissociation
regions.

\section{RRLs and 21-cm Cosmology}\label{sec:cosmo}

There is growing interest in the use of the 21-cm hyperfine transition
of \ion{H}{i} as a cosmological and astrophysical probe of the distant
Universe.  For comprehensive reviews, see \cite{fob06} and
\cite{pl08}.  For a portion of the Universe's history ($1100 \lesssim
z \lesssim 7$), the baryonic content of the intergalactic medium (IGM)
is dominated by \ion{H}{i}, and interactions between the adiabatic
expansion of the IGM and various heating sources (e.g., first stars,
first black holes, dark matter decay) can produce either an absorption
or emission \ion{H}{i} signal relative to the cosmic microwave
background (CMB).

At least three distinct epochs have been identified, which have
corresponding spectral windows for the redshifted \ion{H}{i} signal.
\begin{enumerate}
\item The Dark Ages -- $100 \lesssim z \lesssim 30$ ($15\,\mathrm{MHz}
\lesssim \nu \lesssim 50\,\mathrm{MHz}$).  The \ion{H}{i} gas is expanding and
cooling at a faster rate than the \hbox{CMB}, and collisions within
the gas drive the kinetic temperature of the gas below the CMB
temperature, resulting in an absorption feature.  This epoch is
thought to occur before the first stars form, and, absent any other
heating sources \citep[e.g., energy injection by decaying dark
matter,][]{fop06}, the evolution of this signal should depend only on
cosmological parameters.

\item First star formation -- $30 \lesssim z \lesssim 15$
($50\,\mathrm{MHz} \lesssim \nu \lesssim 90\,\mathrm{MHz}$).  As the first
stars form, they flood the Universe with Lyman-$\alpha$ photons.
These strongly couple the kinetic and spin temperatures of \ion{H}{i}
and produce a second, deep absorption feature.  Detection of this
signal would constrain both the epoch and luminosity function of the
first stars.

\item Epoch of Reionization (EoR) -- $15 \lesssim z \lesssim 7$
($90\,\mathrm{MHz} \lesssim \nu \lesssim 200\,\mathrm{MHz}$).  As stars and the
first black holes heat the gas, its temperature rises above the CMB
temperature.  The absorption feature turns into an emission feature,
which persists until reionization completes and the signal cuts off.
Detection of this feature would constrain the duration of the EoR and
the luminosity function(s) of ionizing sources present.
\end{enumerate}
Estimates of the strength of these \ion{H}{i} signals is both epoch
and model dependent but are in the range 10--100~\mbox{mK}.

One of the key challenges to detecting any of these signals are the
various foregrounds.  These foregrounds include radio frequency
interference (RFI) from terrestrial radio transmitters, ionospheric
phase corruptions, the Galactic synchrotron emission, and emission
from extragalactic sources.  The various spectral windows for these
\ion{H}{i} signals also include frequencies where RRLs have been
observed on Galactic plane sightlines.  It is currently unknown at
what strength RRLs might exist on sightlines at the high Galactic
latitudes where most of the cosmological studies will be made.

{A standard assumption for the removal of \emph{astrophysical}
foregrounds (Galactic synchrotron, extragalactic sources) is that
their spectra are \emph{smooth} \citep{hzb+09}.}  While 21-cm cosmological
observations will clearly be conducted at high Galactic latitudes, in
order to
minimize the contribution of the Galactic synchrotron emission, even
residual effects from RRLs might still vitiate the measurements.

{
As an illustration of the potential impact of RRLs on these
experiments, consider a set of RRLs having an optical depth $\tau \sim
5 \times 10^{-4}$, characteristic of what has been observed to date
(Table~\ref{tab:knownlines}).  We assume no continuum opacity, as is
reasonable for observations at frequencies $\nu \gtrsim 10$~MHz at
high Galactic latitudes.  The continuum temperature~$T_c$ will be the combination of the Galactic synchrotron emission and the
extragalactic background but is likely to be dominated by the former \citep{b67}.  The temperature in the
line is then $(T_c+T)\tau$, for a temperature~$T$ of the gas hosting
the RRLs and employing the usual expansion for small optical depths.
We now consider the two cases, $T_c \gg T$ and $T_c \sim T$.
}

{
Using 150~MHz as a characteristic frequency for detection of the
\ion{H}{i} signal from the \hbox{EoR}, the coldest parts of the sky
have a temperature $T_c \approx 150$~\hbox{K}.  In the Galactic plane, the gas
hosting the RRLs has $T \sim 100$~K (Sect.~\ref{sec:ism}), and if there is
such RRL-hosting gas at high Galactic latitudes, then clearly $T \sim
T_c$.  In this case, spectral fluctuations between RRLs and the continuum of order $(T_c+T)\tau/T_c
\sim 2\tau$ should be expected.  Conversely, if one considers 75~MHz
as a characteristic frequency for the \ion{H}{i} signal from the first
star formation, the coldest parts of the sky have a temperature $T_c
\approx 1000$.  Assuming, as before, $T = 100$~\hbox{K}, spectral fluctuations
of order $\tau$ should be expected.
}

{
The magnitude of the spectral fluctuations expected from the
\ion{H}{i} signal varies with the epoch being considered, but a
typical range of values is $10^{-5}$--$10^{-4}$.  If there is
RRL-hosting gas at high Galactic latitudes, the spectral fluctuations arising
from RRLs could be comparable to or exceed that from the desired
\ion{H}{i} signals.
}

\section{Detection of RRLs}\label{sec:detect}

We begin by reviewing the observational requirements for detecting
RRLs.  From the radiometer equation, the sensitivity of an experiment
designed to detect RRLs can be expressed as
\begin{equation}
\Delta T
 = {T_{\mathrm{sys}}\over{f}}\,\frac{1}{\sqrt{N_{\mathrm{lines}}\,N_{\mathrm{pol}}\,\Delta t_{\mathrm{int}}\Delta\nu}}
\label{eqn:dT}
\end{equation}
where $\Delta T$ is the rms noise, $T_{\mathrm{sys}}$ is the telescope
system temperature, $f$ is the filling factor of the telescope or
array, $\Delta t_{\mathrm{int}}$ is the integration time, $\Delta\nu$ is
the frequency resolution, $N_{\mathrm{pol}}$ is the number of
polarizations, and $N_{\mathrm{lines}}$ is the number of simultaneous transitions observed that can be folded together.

If the RRL has an optical depth~$\tau$, then an ``$m$-sigma'' detection implies
$m\Delta T \le \tau T_{\mathrm{sys}}$, with a corresponding
integration time of
\begin{equation}
\Delta t_{\mathrm{int}}
 = \left(\frac{m}{f\times DF}\frac{T_{\mathrm{sys}}}{\Delta T}\right)^2 {1\over{N_{\mathrm{pol}}\,N_{\mathrm{lines}}\,\Delta \nu}}.
\label{eqn:time}
\end{equation}
{We have introduced the beam dilution factor (dilution factor
or~DF) of the absorbing signal, which is relevant for the case of
observations of lines against a non-discrete background such as the
Galactic synchrotron emission.}
{For example, estimates from two surveys suggest that
absorption lines seen towards the central Galactic plane probably
arise in clouds that subtend between~2\degr\ and~4\degr\ on the sky,
but are unlikely to be bigger \citep{2001JApA...22...51K}.}
If the beam of the antenna is substantially larger than 4\degr, it is
likely that the absorption feature will be diluted by unabsorbed
background radiation.  We estimate DF as the geometric ratio of the
cloud and beam sizes, and $\mathrm{DF} \leq 1$.  The dilution
factor~DF enters equation~(\ref{eqn:time}) as a multiplier of $\Delta
T$, so $\Delta t_{int} \propto \mathrm{DF}^{-2}$.  This factor is not
important when the background source is discrete and thus fully
covered by the absorbing gas, such as in the case of \object{Cas~A}
and other bright Galactic or extragalactic sources.

{%
Most previous RRL observations have been conducted at relatively low
angular resolutions (Table~\ref{tab:previous}).  As a consequence, a
significant uncertainty has been the extent to which different source
regions are blended within the telescope beam.  One of the advantages
of many of the future telescopes (Sect.~\ref{sec:future}) is that they
will have better angular resolution than most of the previous systems
used for RRL observations.Thus, we include the beam dilution
factor~\hbox{DF} as a generic factor to account for a potential
mismatch between the angular sizes of the of foreground absorber(s)
and the background source(s) in the beam.  Indeed, future
observations, particularly those at different frequencies and
therefore different angular resolutions, may help resolve the extent
to which a mismatch has been an issue in previous observations.
}

The system temperature at low radio frequencies is typically dominated
by the synchrotron emission from the Galactic plane itself and thus
varies with both position and frequency.  As can be seen in
Table~\ref{tab:knownlines}, most of the Galactic lines are very weak,
with optical depths $\tau \lesssim 10^{-3}$.  Even a 5$\sigma$
detection (i.e., $m = 5$, equation~\ref{eqn:time}) therefore implies
$\Delta T/T_{\mathrm{sys}} \lesssim 2 \times 10^{-4}$.  Any survey for
these lines should be targeted to reach at least this sensitivity
level.

The lines themselves vary in velocity width from~10 to~100~\kms,
suggesting that a velocity resolution of roughly 2~\kms\ ($\sim
700\,\mathrm{Hz}$ at~100~MHz) is appropriate for blind detection of
the narrowest lines.  However, even observations targeting the broader
lines will benefit from high spectral resolution to aid in the
identification and excision of \hbox{RFI}.  Because the lines
themselves are quite weak, even weak RFI must be carefully removed in
these observations.

\section{Prospects for Future Observations}\label{sec:future}

A new generation of low frequency instruments have either the
opportunity to contribute to Galactic RRL and ISM studies or to be
confused by the lines when doing other studies
(Table~\ref{tab:future}).  Improvements in both sensitivity and
resolution will help further resolve issues of physical cloud
parameters such as size, pressure and temperature, while allowing a
more complete census of the ionized carbon gas in the Galaxy.
Moreover, in many cases, one of the prime science drivers for the
telescope is observations of the cosmological \ion{H}{i} signal, so
that careful removal of foregrounds, including RRLs, is paramount.  We
shall consider the performance of these telescopes in turn.

{At low frequencies, the system temperature~$T_{\mathrm{sys}}$
  should be dominated by the Galactic synchrontron emission, and all
  of the telescopes will have the capability to conduct dual
  polarization observations ($N_{\mathrm{pol}} = 2$).}
{Thus, with respect to the performance of these future
telescopes, most notably those still in the design phase,
equation~\ref{eqn:time} indicates that there are only two other
factors under the control of those designing the telescope.}
One is the filling factor~$f$ of the telescope and the other is the
number of lines~$N_\mathrm{lines}$, which is coupled to the total
bandwidth that the system can handle.

{One feature common to all of these future systems is the
signal chain will be largely digital, which should enable significant
improvements in the stability and shape of the spectral bandpasses.}
{Further, all of the telescopes will have a significantly
  enhanced frequency tuning range (with typical ranges of~2:1 or
  better), which will enable many more lines to be observed.}
{Finally, all of the systems that we will discuss will be dipole-based
phased array systems in which the dipoles will be deployed at
essentially ground level.}
Such systems are in contrast to single dish observations
in which the feed is at an altitude of tens of meters or more.  At
such altitudes, a feed is susceptible to many low-level, scattered RFI
sources, including those from over the horizon, and experience has
shown that single dish RRL observations are rarely, if ever,
successful during the day, even on days when the general RFI level
should be quite low (e.g., the Easter holiday).  Dipoles deployed at
ground level should suppress RFI sources located near the horizon and
should enable more sensitive RRL observations,\footnote{
Presumably the suppression of RFI should be less important for the
Lunar Radio Array (Sect.~\ref{sec:lra}).}
a hypothesis generally confirmed by daytime observations with the
dipole-based UTR-2.

\subsection{Long Wavelength Array (LWA)}\label{sec:lwa1}

The Long Wavelength Array (LWA) is intended to be a next-generation,
high angular resolution imaging telescope operating at frequencies
between~20 and~80~MHz \citep{ellingson09}.  Some of the key science
projects are the high redshift Universe, including searches for
high-redshift radio galaxies and studies of clusters of galaxies;
probing particle acceleration in supernova remnants, radio galaxies,
and clusters of galaxies; and searches for radio transients.  The
telescope would be composed of as many as 50 dipole phased array
``stations,'' with each station containing 256 dual polarization
dipoles.  The first station (LWA-1) is currently under construction
and is planned to conduct science observations in its own right.

The configuration of LWA-1 is a pseudo-random distribution of the
dipoles \citep{kc09}, aimed at minimizing the sidelobes from the
phased-array station.  Because of the broad frequency range, the
filling factor of the station is frequency dependent.

Table~\ref{tab:lwadetect} summarizes our estimates for the required
integration times to detect Galactic plane RRLs with LWA-1 over
its operational frequency range.  We assume $5\sigma$ detections of
roughly $10^{-3}$ optical depth lines at roughly 2~\kms\ velocity
resolution.
{We have also assumed an absorber size of 4\degr, which is
comparable to the station resolution at the higher frequencies.}


This first station of the planned LWA instrument is ideal for RRL
detection work due to both its flexible frequency coverage and its
high filling factor at lower frequencies.  At optimum frequencies, the
desired sensitivity may be reached in as little as 6~hours.  

While there has been some consideration of precursor Dark Ages
observations with LWA-1, the prime science mission for the
LWA as a whole does not include cosmological \ion{H}{i} measurements.
Moreover, it is not clear that LWA-1 would have the requisite
sensitivity to detect the Dark Ages signal.  Consequently, RRLs are
not likely to be important as a contaminant in any other LWA-1
observations.

\subsection{LOw Frequency ARray (LOFAR)}\label{sec:lofar}

The {LOw Frequency ARray (LOFAR)}\footnote{
http://www.lofar.org
} is a next-generation, low radio frequency telescope operating at
frequencies between~30 and~240~MHz.  Its frequency coverage is split
between a high-band antenna (\hbox{HBA}, 110--240~MHz) and low-band
antenna (\hbox{LBA}, [10]~30--90~MHz).

One of its key science projects is to study the Epoch of Reionization
using the high-band antenna at redshifts $11.5> z > 6.5$ ($115\,\mathrm{MHz}
< \nu < 190\,\mathrm{MHz}$).  This is the frequency range over which the RRLs
switch from absorption to emission; at some frequencies during this
transition they may be undetected \citep{1989ApJ...341..890P}.  

In its full configuration, LOFAR has a very low areal filling factor,
making it unsuitable for RRL detection and study.  However using the
``super-station'' core of 6 stations within a 300~m diameter area or
using a single station, more reasonable filling factors of order 10\%
can be achieved.
{The instrument has flexible frequency coverage and
it is possible to cover large bandwidths (up to~48~MHz) simultaneously.}
The ability to fold together many lines will help with detection.
{However, the frequency setup has a fixed frequency resolution,
corresponding to about~5~\kms\ velocity resolution at the lowest
frequencies; thus the narrower RRLs may be under-resolved.}

By folding together hundreds of transitions and using the
super-station core, this instrument would be able to detect $10^{-3}$
optical depth lines in a few hours at the lowest frequencies; longer
integrations should be able to detect these and fainter lines at
higher frequencies or with less folding or both.
{Folding together adjacent RRLs assumes that $\Delta n \ll n$,
 and thus that the transitions are essentially identical.}
{This assumption is unlikely to be warranted if hundreds of
 lines are folded together, and the resulting line profile would no
 longer be useful for deriving physical parameters of the absorbing
 gas.}
The planned EoR experiment will
integrate deeply enough to detect RRLs, if present, at the higher
frequencies.

\subsection{Square Kilometre Array (SKA)}\label{sec:ska}

The {Square Kilometre Array (SKA)}\footnote{
http://www.skatelescope.org
} will be one of a suite of new, large astrophysics facilities for the
$21^{\mathrm{st}}$ century and will probe fundamental physics, the origin
and evolution of the Universe, the structure of the Milky Way Galaxy,
and the formation and distribution of planets.  
{Highlighting the SKA Science Case are Key Science Projects
\citep[KSPs,][]{cr04,g04c}, one of which is ``Probing the Dark Ages
and the Epoch of Reionization'' \citep{cfbjrf04}.}
The goal is to track the transition of the originally neutral IGM into
its current mostly ionized state via imaging of the
(highly-redshifted) \ion{H}{i} line.  With reference to the three
21-cm cosmology epochs (Sect.~\ref{sec:cosmo}), the goal for the SKA is to
cover the \hbox{EoR}, and as far back into the epoch of first star
formation as possible ($20 \lesssim z \lesssim 7$, $70\,\mathrm{MHz}
\lesssim \nu \lesssim 200\,\mathrm{MHz}$).

The configuration of the low-frequency component of the \hbox{SKA},
the so-called SKA-lo, remains a matter of
active study, but the current specifications forsee approximately 25\%
of the total collecting area within a diameter of~1~km and
approximately 50\% of the total collecting area within a diameter
of~5~km \citep{ska100}.  The resulting filling factors are $f \sim
0.5$ ($f \sim 0.05$) within the central 1~km (5~km).

Scaling from the results in Table~\ref{tab:lwadetect}, the central
portion of the SKA-lo could therefore detect RRLs along the Galactic
plane in integration times of order 5~hr.  Given the primary
scientific driver for the SKA-lo, however, it is unlikely that it will
be conducting deep observations along the Galactic plane.  Rather, it
will likely conduct its deep observations ($\gtrsim 100$~hr) at high
Galactic latitudes.
{The strength and distribution of RRLs at high Galactic
latitudes is currently unstudied; it is unknown if they will be
present at all.}
{However, even if the RRLs on these sightlines are an order of
  magnitude weaker than currently known lines, with $\tau \sim
  10^{-4}$, equation~(\ref{eqn:time}) shows that they would still be
  detected in integrations of order 100~hours.}

\subsection{Lunar Radio Array (LRA)}\label{sec:lra}

{The Lunar Radio Array (LRA) is a concept for a radio telescope
operating at frequencies around~100~MHz and sited on the far side of
the Moon with a prime mission of making precision 21-cm cosmological
measurements.}
With reference to the three 21-cm cosmology epochs (\ref{sec:cosmo}),
the goal for the LRA is to probe at least the epoch of first star
formation, and as deeply into the Dark Ages as possible ($100 \lesssim
z \lesssim 15$, $10\,\mathrm{MHz} \lesssim \nu \lesssim
100\,\mathrm{MHz}$).

Some of the initial designs for the LRA follow the configuration
analysis of \cite{lzmzh08}, who advocate a highly compact ``super
core.''  For such a configuration, the filling factor is large,
potentially as large as $f \sim 0.9$.  The result would be to reduce
the required integration times potentially by an order of magnitude.
However, like the \hbox{SKA-lo}, the LRA would most likely target fields
well away from the Galactic plane.  Similar comments apply to the LRA
as to the \hbox{SKA-lo} regarding such observations.

\section{Conclusions}\label{sec:conclude}

{
We have summarized the state of current observations and the potential
for future observations of decametric-wavelength radio recombination
lines (RRLs, $\nu \lesssim 200$~MHz).  These lines are of interest
both because they offer a sensitive and unique probe of cooler, less
dense ionized regions throughout the Galaxy than is probed by
traditional higher frequency RRL studies, as well as potentially
presenting a confusing foreground for redshifted 21-cm
cosmology and astrophysical measurements from the Epoch of
Reionization (EoR) and Dark Ages.
}

There are a number of low radio frequency arrays either under
construction or in design and development, including the Long
Wavelength Array (LWA), the LOw Frequency ARray (LOFAR), the
low-frequency component of the Square Kilometre Array (SKA-lo), and
the Lunar Radio Array (LRA).
{All of these arrays will be capable of observing at
frequencies below about~200~MHz, and they could be used to study RRLs,
but the most important design factor for a telescope is the areal
\emph{filling factor}.}
{For the current arrays, LWA-1 has a filling factor that varies
with frequency, from nearly 100\% at its lowest frequencies ($\approx
20$~MHz) to around 10\% at the highest frequencies ($\approx 80$~MHz);
the super-station core of LOFAR ranges from~0.3 to~0.02 over a similar
frequency range (its LBA).}  
{The estimated integration times are 10 to~100~hr to detect
RRLs at optical depths of order $10^{-3}$; in the case of the LOFAR
super-station core, these relatively modest integration times are
achieved in part by folding potentially hundreds of lines over large
bandwidths (up to~48~MHz).}
The SKA-lo and the LRA are still in the design phase, but might likely
have higher filling factors, potentially larger than 50\%, leading to
the detection of these lines in 10~hr or less.
{Further, all of these telescopes offer the possibility of
  improved performance as a function of frequency, including both improved
  spectral bandpasses and larger frequency tuning ranges.}

RRLs have been studied extensively along the line of sight to
the bright radio source \object{Cas~A}.  However, the weakness of
the lines and the subsequent need for long integration times and
careful data processing to remove even very low level radio frequency
interference (RFI) have hampered efforts for the type of systematic,
multi-frequency surveys required to probe the physics of the cool, low
density ISM throughout the Galaxy.
{Although not a key science driver for any of these telescopes,
systematic surveys for RRLs could be important secondary science to
emerge.}

{
From the standpoint of 21-cm studies of the Epoch of Reionization and
the Dark Ages, deep observations ($> 100$~hr) will likely be conducted
with many of these telescopes to search for the highly redshifted
\ion{H}{i} signatures, implying sensitivity to much lower optical depths than has heretofore been
possible, of order $10^{-4}$.  Simple estimates suggest that, if there
is RRL-hosting gas at the high Galactic latitudes that will be
targeted for the observations, the RRLs could impose spectral
fluctuations that are comparable to or exceed the expected \ion{H}{i}
signatures.  One significant mitigating factor, however, is that the
lines are likely to be quite narrow, relative to the expected
\ion{H}{i} features.  \ion{H}{i} signatures from the EoR or Dark Ages
are expected to have widths of order 0.2~MHz or larger; by constrast,
a velocity width of order 10~\kms\ near~100~MHz implies a line width
of only a few kHz.  Thus, the RRLs might likely be narrow, with a low 
spectral occupancy.  The lines might easily be treated as
non-terrestrial \hbox{RFI}.
}

{%
Two final comments are warranted regarding extremely deep
observations, particularly as they relate to efforts to detect the
\ion{H}{i} signal from the Epoch of Reionization and before.  It is
not yet known what is the most relevant frequency range for 21-cm
cosmology.  However, it is possible that LOFAR-HBA and SKA-lo, and
possibly the future \hbox{LRA}, would be observing in the regime for
which RRLs transition from emission at higher frequencies to
absorption at lower frequencies.  In this case, the lines would not be
detected.  Second, most of our comments have concerned
$\alpha$-transition lines.  In deep integrations, $\beta$-, $\gamma$-,
and potentially $\delta$-transition lines can be detected
\citep{2007MNRAS.374..852S}.  The resulting spectral occupancy would
be much higher and the ability to fold or remove the lines could be
reduced substantially.
}

\begin{acknowledgements}  
{The authors thank E.~Polisensky for providing sky
temperatures, A.~Cohen for help with calculating the station filling
factor for the LWA-1, G.~Smirnov for information about previous instruments,
and the referee for suggestions that helped us
clarify and quantify certain points.}
This research has made use of NASA's Astrophysics Data
System.
The {LUNAR consortium} is funded by the NASA Lunar Science 
  Institute (via Cooperative Agreement NNA09DB30A) to investigate 
  concepts for astrophysical observatories on the Moon. 
Basic research in radio astronomy at the Naval Research Laboratory is funded
by 6.1 Base funding.
\end{acknowledgements}

\clearpage
\onecolumn

\begin{center}
\tablecaption{Known Galactic Carbon $\alpha$ RRL Absorption
  Lines\tablefootmark{a}\label{tab:knownlines}}
\tablehead{%
\hline\hline
{Sightline} & {$\nu$} & {$n$} & {$\tau$}             & {$V_{\mathrm{LSR}}$}
   & {$\Delta v$} & {Reference} \\
            & {(MHz)} &       & {($\times 10^{-3}$)} & {(\kms)}             
   & {(\kms)}     & \\
\hline}
\tabletail{\hline}
\begin{supertabular}{lcccccl}
\\
\protect\object{G0$+$0}   & 34.5  & 575 & 1.16 & 5.3$\pm$0.8 & 20.5$\pm$1.1 & 5 \\
                          & 75    & 443 & 0.57$\pm$0.04 & $-1.0\pm$0.5 & 14.8$\pm$1.2 & APE88 \\
                          & 76    & 441 & 0.73 & $-1$ & 24 & 1 \\

\protect\object{G0$-$4}\tablefootmark{b}
                          & 76   & 441 & 0.33 &  $-6$ & 13 & 1 \\
\protect\object{G0$-$2}   & 76   & 441 & 0.57 & $-10$ & 30 & 1 \\
\protect\object{G0$+$0}   & 34.5 & 575 & 1.16 &  5.3$\pm$0.8 & 20.5$\pm$1.1 & 5 \\
                          & 76 	 & 441 & 0.73 & $-1$ & 24 & 1 \\
\protect\object{G0$+$2}   & 76 	 & 441 & 0.70 & $-2$ & 26 & 1 \\
\protect\object{G0$+$4}   & 76 	 & 441 & 0.68 & $-2$ & 31 & 1 \\
\protect\object{G2$-$3.5}\tablefootmark{b}
                          & 76 	 & 441 & 0.52 &    7 & 5 & 1 \\
\protect\object{G2$-$2}   & 76 	 & 441 & 0.97 &    5 & 9 & 1 \\
\protect\object{G2$+$0}   & 76 	 & 441 & 0.90 &    2 & 25 & 1 \\
\protect\object{G2$+$2}   & 76 	 & 441 & 0.61 &    6 & 11 & 1 \\
\protect\object{G3$+$0}   & 76 	 & 441 & 0.90 & $-1$ & 14 & 1 \\
\protect\object{G4$+$0}   & 76 	 & 441 & 0.54 &    8 & 17 & 1 \\
\protect\object{G5$+$0}   & 34.5  & 575 & 0.74 & 10.2$\pm$1.4 & 21.2$\pm$2.0 & 5 \\
\protect\object{G6$+$0}   & 76    & 441 & 0.73 &    9 & 25 & 1 \\
\protect\object{G6.6$-$0.2} & 76  & 441 & 0.87 &   10 & 28 & 1 \\
\protect\object{G8$+$0}      & 76    & 441 & 1.09 &   11 & 22 & 1 \\
\protect\object{G10$+$0}     & 34.5  & 575 & 0.81 & 14.3$\pm$1.7 & 37.5$\pm$2.5 & 5 \\
                             & 	    &     & 0.93 &  9.6$\pm$1.5 & 24.9$\pm$2.3 & 5 \\
                             & 	    &     & 0.47 & 40.2$\pm$2.6 & 18.4$\pm$3.7 & 5 \\
                             & 76   & 441 & 0.72 &   17 & 26 & 1 \\
\protect\object{G12$+$0.0}   & 76   & 441 & 0.91 &   12 & 20 & 1 \\
\protect\object{G14$-$2.0}   & 76   & 441 & 0.88 &   11 & 32 & 1 \\   
\protect\object{G14$+$0}     & 34.5 & 575 & 0.66 & 37.8$\pm$2.7 & 54.0$\pm$3.8 & 5 \\
                             & 76   & 441 & 0.85 &   16 & 25 & 1 \\
\protect\object{G14$+$2.0}   & 76   & 441 & 0.51 &   21 & 28 & 1 \\
\protect\object{G16$+$0.0}   & 76   & 441 & 0.76 &   14 & 47 & 1 \\
\protect\object{M16}         & 68   & 456 & 1.8$\pm$0.3 & 18$\pm$2 & 16$\pm$3 & 6 \\
                             & 80   & 435 & 2.0$\pm$0.3 & 20$\pm$1 & 13$\pm$2 & 6 \\
\protect\object{G16.5$+$0}   & 34.5 & 575 & 0.59 & 26.4$\pm$2 & 32.8$\pm$2.8 & 5 \\
\protect\object{G06.9$+$0.8} & 76   & 441 & 1.07 &   20 & 21 & 1 \\
\protect\object{G18$+$0}     & 76   & 441 & 0.77 &   15 & 33$\pm$6 & 1 \\
\protect\object{G20$+$0}\tablefootmark{b}  
                             & 76   & 441 & 0.59 &   36 & 18$\pm$3 & 1 \\
\protect\object{G63$+$0}     & 34.5 & 575 & 0.42 & 36.2$\pm$3.2 & 46.0$\pm$4.5 & 5 \\
\protect\object{G75$+$0}     & 25   & 640 & 1    &   12 & 15 & 4 \\ 
\protect\object{G75$+$0}     & 34.5 & 575 & 0.40 & 6.9$\pm$1.8 & 27.1$\pm$2.6 & 5 \\
\protect\object{DR~21}       & 25   & 640 & 0.7$\pm$0.3 & 0$\pm$8 & 42$\pm$12 & 2 \\
                             & 34.5 & 575 & 0.74 & 4.5$\pm$1.6 & 18.8$\pm$2.2 & 5 \\
\protect\object{S140}        & 25   & 640 & 1$\pm$0.3 & $-6\pm$17 & 96$\pm$24 & 2 \\
                             &      &     & 0.5$\pm$0.3 & $-36\pm$17 & 96$\pm$24 & 2 \\
\protect\object{NGC~2024}    & 25   & 640 & 1    &  10  & 36 & 4 \\
\protect\object{G287.4$-$0.6}\tablefootmark{b} 
                             & 76 & 441 & 0.48 & $-23$ & 38 & 1 \\
\protect\object{G312$+$0}    & 76 & 441 & 0.88 & $-53$ & 16 & 1 \\
\protect\object{G340$+$0}\tablefootmark{b} 
                             & 76 & 441 & 0.27 & $-39$ & 30 & 1 \\
\protect\object{G342$+$0}\tablefootmark{b}
                             & 76 & 441 & 0.53 & $-39$ & 17 & 1 \\
\protect\object{G344$+$0}    & 76 & 441 & 0.65 & $-27$ & 26 & 1 \\
\protect\object{G346$+$0}\tablefootmark{b} 
                             & 76 & 441 & 0.20 & $-20$ & 20 & 1 \\
\protect\object{G348$+$0}    & 76 & 441 & 0.53 & $-12$ & 14 & 1 \\
\protect\object{G350$+$0} & 76 	 & 441 & 0.81 & $-11$ & 14 & 1 \\
\protect\object{G352$-$2.0} & 87  & 441 & 0.70 & $-17$ & 23 & 1 \\
\protect\object{G352$+$0} & 34.5  & 575 &  0.65 & -1.2$\pm$1.8 &  34.8$\pm$2.6 & 5 \\
         & 76    & 441 & 1.25 & $-10$ & 11 & 1 \\
\protect\object{G352$+$2.0} & 76  & 441 & 0.83 &  $-8$ & 22 & 1 \\
\protect\object{G354$+$0} & 76 	 & 441 & 0.87 & $-10$ & 26 & 1 \\
\protect\object{G356$+$0} & 76 	 & 441 & 0.76 &  $-4$ & 17 & 1 \\
\protect\object{G358$-$2} & 76 	 & 441 & 0.53 & $-10$ & 28 & 1 \\
\protect\object{G358$+$0} & 76 	 & 441 & 0.90 &  $-3$ & 24 & 1 \\
\protect\object{G358$+$2} & 76 	 & 441 & 0.80 &     3 &  8 & 1 \\ 
L1407    & 25    & 640 & 0.7         & $-10$      & 17        & 3 \\
\end{supertabular}

\tablefoottext{a}{This listing excludes lines for the \protect\object{Cas~A} line of sight.}
\tablefoottext{b}{Tentative detection}
\tablebib{%
  (1)~\citet{1995ApJ...454..125E};
  (2)~\citet{1991SvAL...17....7G}; 
  (3)~\citet{1991SvAL...17...10G};
  (4)~\citet{1984SvAL...10..384K};
  (5)~\citet{2001JApA...22...51K};
  (6)~\citet{1988MNRAS.235..151A}.
}  
\end{center}

\clearpage

\twocolumn
{
\begin{table}
\caption{Previous and Existing Instrument
  Capabilities}\label{tab:previous}
\centering
\begin{tabular}{lcccc}
\hline\hline
Name     & Frequency & Resolution  & filling factor & $T_{\mathrm{detect}}$ \\
         & (MHz)     &             &                & (hr) \\
\hline
UTR-2    & 26        & 40\arcmin   &  1             &   1 \\
Gauribidanur & 34.5  & $21\arcmin \times 25\degr$
                                   & 0.59           & 1.5 \\
DKR-1000 & 40        & $44\arcmin \times 1\degr$    
                                   & 0.2            &  25 \\
Arecibo  & 47        & 80\arcmin   & 0.25           & 23 \\
Parkes   & 76        & 4\degr      & 0.25           & 2 \\
\hline
\end{tabular}
\tablefoot{$T_{\mathrm{detect}}$ is the time to detect (5$\sigma$) a line of
  optical depth $10^{-3}$ with an well-matched instrumental bandwidth.
  The instruments listed are ones from which published RRL
  results have been obtained.  In many cases, characteristics have
  been taken from the literature, and the instrument itself may not be
  currently operational.}
\end{table}
}

\begin{table}
 \caption{Future Telescope Capabilities\label{tab:future}}
 \centering
 \begin{tabular}{lcccc}
\hline\hline
Name     & Frequency & Resolution  & filling factor & $T_{\mathrm{detect}}$ \\
         & (MHz)     &             &                & (hr) \\
\hline
LWA-1    & 20--80    & 9\degr--2\degr & 0.9--0.1    & $\sim 10$ \\
LOFAR-LBA & 30--90    & 2\degr--0.6\degr & $\sim 0.1$ & $\sim 10$ \\
LOFAR-HBA & 110--240  & 30\arcmin--15\arcmin & $\sim 0.1$ & ... \\
SKA-lo    & 150      & 7\arcmin       & $\sim 0.1$  & $< 10$ \\
LRA       & 100      & 10\arcmin      & $\sim 0.9$  & $< 10$ \\
\hline
 \end{tabular}
 \tablefoot{Values listed are characteristic or indicative, but
    none of these telescopes have entered an operational state, so
    actual performance may vary.  For \hbox{LOFAR-LBA}, bandwidths of~48~MHz have
    been assumed.  \hbox{LOFAR-HBA} covers a poorly explored frequency range
    in which the lines may become undetectable as they transition from
    emission at higher frequencies to absorption at lower frequencies;
    the same may also be true of SKA-lo.  Particularly for the SKA-lo and
    \hbox{LRA}, final design parameters are yet to be determined, and
    these values are indicative.}
\end{table}

\begin{table}
\caption{Estimates for Detecting Galactic Plane RRLs with
	LWA-1}\label{tab:lwadetect}
\centering
\begin{tabular}{ccccccc}
\hline\hline
{$\nu$} & {$n$} & {$f$} & {$N_{\mathrm{lines}}$} & DF 
	& {$T_{\mathrm{sys}}$} 
	& {$\Delta t_{\mathrm{int}}$} \\
{(MHz)} &       &	&                        &
	& {($\times 10^4$~K)} 
	& {(hr)} \\
\hline
20 & 683 & 0.92 & 10 & 0.26 &  29.5 & 50 \\
25 & 640 & 0.87 &  8 & 0.41 &  18.3 & 28 \\ 
30 & 603 & 0.76 & 12 & 0.59 &  12.4 & 6 \\
35 & 572 & 0.63 & 10 & 0.8  &   8.9 & 5.5 \\
40 & 548 & 0.5  &  8 & 1    &   6.7 & 7 \\
45 & 526 & 0.4  &  6 & 1    &   5.2 & 15 \\
50 & 508 & 0.33 &  6 & 1    &   4.1 & 23 \\
55 & 492 & 0.27 & 10 & 1    &   3.3 & 10 \\
60 & 478 & 0.23 & 10 & 1    &   2.7 & 14 \\
65 & 466 & 0.2  &  8 & 1    &   2.3 & 24 \\
70 & 454 & 0.17 &  8 & 1    &   1.9 & 32 \\
75 & 444 & 0.15 &  6 & 1    &   1.7 & 56 \\
80 & 434 & 0.13 &  6 & 1    &   1.4 & 72 \\
\hline
\end{tabular}
\tablefoot{$T_{\mathrm{sys}}$ is derived from the GSM sky models
\protect\citep{2008MNRAS.388..247D}.  It is the average temperature along
the Galactic Plane towards Galactic center ($-5\degr < b < 5\degr$, $340\degr\ < \ell < 20\degr$) after convolving the models to
LWA-1 beam size and removing the pixels with the highest and lowest 2\% of intensity.  The
dilution factor, \hbox{DF}, assumes 4\degr\ cloud sizes.  The filling factor~$f$ is derived using the station dipole layout design \protect\citep{kc09} and
accounts for overlaps and edge effects.  $N_{\mathrm{lines}}$ is based
on two contiguous bands with roughly 2~\kms\ velocity
resolution.  The integration time to detect lines with
optical depths of $10^{-4}$ is 100$\times$ the values listed here.} 
\end{table}

\clearpage


\clearpage

\begin{thebibliography}{}

\bibitem[\protect\citeauthoryear{Anantharamaiah et al.}{1988}]{1988MNRAS.235..151A}
         Anantharamaiah, K.~R., Payne, H.~E., \& Erickson, W.~C.  1988, \mnras, 235, 151

\bibitem[\protect\citeauthoryear{Blake et al.}{1980}]{1980Natur.287..707B}
	Blake, D.~H., Crutcher, R.~M., \& Watson, W.~D.\ 1980, \nat,
	287, 707

\bibitem[\protect\citeauthoryear{Bridle}{1967}]{b67}
        Bridle, A.~H.  1967,
	\mnras, 136, 219

\bibitem[\protect\citeauthoryear{Carilli et al.}{2004}]{cfbjrf04}
	Carilli, C.~L., Furlanetto, S., Briggs, F., Jarvis, M.,
	Rawlings, S., \& Falcke, H.  2004, 
	in Science with the Square Kilometre Array, eds.\
	C.~L.~Carilli \& S.~Rawlings (Elsevier: Amsterdam) p.~1029

\bibitem[\protect\citeauthoryear{Carilli \& Rawlings}{2004}]{cr04}
	Carilli, C.~L., \& Rawlings, S.  2004, \textit{Science with
	the Square Kilometre Array}, New Astron.\ Rev.\ (Elsevier:
	Amsterdam)

\bibitem[\protect\citeauthoryear{Ellingson et al.}{2009}]{ellingson09}
	Ellingson, S., Clarke, T.~E., Cohen, A., Craig, J., Kassim,
	N.~E., Pihlstrom, Y., Rickard, L.~\hbox{J}, \& Taylor, G.~B.  2009, 
	Proc.\ \hbox{IEEE}, 97, 1421-1430

\bibitem[\protect\citeauthoryear{Erickson et al.}{1995}]{1995ApJ...454..125E}
	Erickson, W.~C., McConnell, D., \& Anantharamaiah, K.~R.\
	1995, \apj, 454, 125

\bibitem[\protect\citeauthoryear{Furlanetto, Oh, \& Briggs}{Furlanetto et al.}{2006}]{fob06}
	{Furlanetto}, S.~R., {Oh}, S.~P., \& {Briggs}, F.~H.  2006,
	Phys.\ Reports, 433, 181

\bibitem[\protect\citeauthoryear{Furlanetto, Oh, \& Pierpaoli}{Furlanetto et al.}{2006}]{fop06}
	{Furlanetto}, S.~R., {Oh}, S.~P., {Pierpaoli}, E.  2006,
	\prd, 74, 103502

\bibitem[\protect\citeauthoryear{Gaensler}{2004}]{g04c}
	Gaensler, B.{}M.  2004, ``Key Science Projects for the SKA''
	SKA Memorandum~44, {http://www.skatelescope.org}

\bibitem[\protect\citeauthoryear{Golynkin \& Konovalenko}{1991a}]{1991SvAL...17....7G}
	Golynkin, A.~A., \& Konovalenko, A.~A.  1991a, Soviet Astron.\
	Lett., 17, 7

\bibitem[\protect\citeauthoryear{Golynkin \& Konovalenko}{1991b}]{1991SvAL...17...10G}
	Golynkin, A.~A., \& Konovalenko, A.~A.  1991b, Soviet Astron.\ Lett., 17, 10

\bibitem[\protect\citeauthoryear{Gordon \& Sorochenko}{2002}]{gs02}
        Gordon, M.~A., \& Sorochenko, R.~L.  2002,
	Radio Recombination Lines: Their Physics and Astronomical
	Applications (Kluwer: Dordrecht)

\bibitem[\protect\citeauthoryear{Harker et al.}{2009}]{hzb+09}
        Harker, G., Zaroubi, S., Bernardi, G., et al.  2009, 
	\mnras, 397, 1138

\bibitem[\protect\citeauthoryear{Helmboldt \& Kassim}{2009}]{hk09}
	Helmboldt, J., \& Kassim, N.~E.  2009, \aj, 138, 838

\bibitem[\protect\citeauthoryear{Kantharia \& Anantharamaiah}{2001}]{2001JApA...22...51K}
	Kantharia, N.~G., \& Anantharamaiah, K.~R.\ 2001, J.\ Astrophys.\ Astron., 22, 51

\bibitem[\protect\citeauthoryear{Kantharia et al.}{1998}]{1998ApJ...506..758K}
	Kantharia, N.~G., Anantharamaiah, K.~R., \& Payne, H.~E.\
	1998, \apj, 506, 758

\bibitem[\protect\citeauthoryear{Kogan \& Cohen}{2009}]{kc09}
	Kogan, L., \& Cohen, A.  2009, 
	LWA Memorandum~\#150; {http://www.ece.vt.edu/swe/lwa/lwa0150.pdf}

\bibitem[\protect\citeauthoryear{Konovalenko}{1984}]{1984SvAL...10..384K}
	Konovalenko, A.~A.  1984, Soviet Astron.\ Lett., 10, 384 

\bibitem[\protect\citeauthoryear{Konovalenko \& Sodin}{1980}]{1980Natur.283..360K}
	Konovalenko, A.~A., \& Sodin, L.~G.\ 1980, \nat, 283, 360


\bibitem[\protect\citeauthoryear{Lidz et al.}{2008}]{lzmzh08}
	Lidz, A., Zahn, O., McQuinn, M., Zaldarriaga, M., \&
	Hernquist, L.  2008, 
	\apj, 680, 962

\bibitem[\protect\citeauthoryear{de Oliveira-Costa et al.}{2008}]{2008MNRAS.388..247D} 
        de Oliveira-Costa, A., Tegmark, M., Gaensler, B.~M., Jonas, J., 
        Landecker, T.~L., \& Reich, P.\ 2008, \mnras, 388, 247 


\bibitem[\protect\citeauthoryear{Payne et al.}{1994}]{1994ApJ...430..690P}
	Payne, H.~E., Anantharamaiah, K.~R., \& Erickson, W.~C.\ 1994,
	\apj, 430, 690

\bibitem[\protect\citeauthoryear{Payne et al.}{1989}]{1989ApJ...341..890P}
	Payne, H.~E., Anantharamaiah, K.~R., \& Erickson, W.~C.\ 1989,
	\apj, 341, 890

\bibitem[\protect\citeauthoryear{Pritchard \& Loeb}{2008}]{pl08}
	Pritchard, J.~R., \& {Loeb}, A.  2008, \prd, 78, 103511 

\bibitem[\protect\citeauthoryear{Schilizzi et al.}{2007}]{ska100}
	Schilizzi, R.~T., et al.  2007, 
	``Preliminary Specifications for the Square Kilometre Array,''
	SKA Memorandum~100; 
	\texttt{http://www.skatelescope.org/PDF/memos/100\_Memo\_Schilizzi.pdf}

\bibitem[\protect\citeauthoryear{Sorochenko \& Smirnov}{1990}]{1990ASSL..163..189S}
	Sorochenko, R.~L., \& Smirnov, G.~T.\ 1990, IAU Colloq.~125:
	Radio Recombination Lines: 25 Years of Investigation, 163, 189

\bibitem[\protect\citeauthoryear{Stepkin et al.}{2007}]{2007MNRAS.374..852S}
	Stepkin, S.~V., Konovalenko, A.~A., Kantharia, N.~G., \& Udaya
	Shankar, N.\ 2007, \mnras, 374, 852
\end{thebibliography}
\end{document}